\documentclass{PoS}

\def\be{\begin{eqnarray}}
\def\ee{\end{eqnarray}}
\def\mhalf{m_{1/2}}

\newcommand{\lsim}{\raisebox{-0.13cm}{~\shortstack{$<$ \\[-0.07cm] $\sim$}}~}
\newcommand{\gsim}{\raisebox{-0.13cm}{~\shortstack{$>$ \\[-0.07cm] $\sim$}}~}

\title{Revisiting No-Scale Supergravity Inspired Scenarios}

\ShortTitle{No-Scale Supergravity}

\author{\speaker{Amine Benhenni}, Jean-Lo\"{i}c Kneur, Gilbert Moultaka,\\
        Universit\'e Montpellier 2, Laboratoire Charles Coulomb UMR 5221, F-34095, Montpellier, France\\
  E-mail: \email{amine-lies.benhenni@um2.fr}, \email{jean-loic.kneur@um2.fr}, \email{gilbert.moultaka@um2.fr}}

\author{ Sean Bailly\\
       Laboratoire de Physique Th\'eorique LAPTH, Universit\'e de Savoie, CNRS (UMR 5108)\\
9 chemin de Bellevue, BP 110, F-74941 Annecy-Le-Vieux Cedex, France\\
       E-mail: \email{sean.bailly@lapth.in2p3.fr}}

\abstract{We consider no-scale supergravity inspired scenarios, emphasizing the possible 
dynamical determination of the soft supersymmetry-breaking parameters as triggered by the radiative 
corrections that lift an essentially flat tree-level potential in the hidden sector. 
We (re)emphasize the important role played by the scale-dependent vacuum energy contribution to the 
effective potential for the occurrence of consistent no-scale minima. The most relevant input parameters 
are introduced as $B_0$ (the soft breaking mixing Higgs parameter) and $\eta_0$ 
(the cosmological constant value at high energy) instead of $\mhalf$ and $\tan \beta$, 
the latter being determined through a (generalized) potential minimization at electroweak scales. 
We examine the theoretical and phenomenological viability of such a mechanism when confronted with 
up-to-date calculations of the low energy sparticle spectrum and with present constraints
from the LHC and other observables. The tight dark matter relic density constraint for a neutralino LSP
scenario can be considerably relaxed for a gravitino LSP scenario possible in this framework.}

\FullConference{The 2011 Europhysics Conference on High Energy Physics, EPS-HEP 2011,\\
		July 21-27, 2011\\
		Grenoble, Rh\^one-Alpes, France}

\begin{document}

\section{Introduction}
No-scale supergravity models~\cite{Cremmer:1983bf,Ellis:1983sf,Ellis:1983ei} 
are a specific set of supergravity models, 
in which the vanishing of the tree-level potential in the
hidden sector direction can be automatic for an appropriately chosen form of the K\"ahler potential.
 Moreover,  the value of the gravitino mass     
$m_{3/2}$ can be fixed dynamically by (non-gravitational) radiative correction stabilization, and is related 
to other soft SUSY-breaking parameters.
This no-scale mechanism has been known for a long time, but the  
complexity of a full minimization of the effective potential lead in the early days to consider only 
specific approximations. More recently the strict no-scale boundary conditions $m_0=A_0=0$ 
have often been studied for their phenomenological consequences but without specifying a precise link
with the above-mentioned scalar potential minimization.
Modern MSSM spectrum calculation tools allow
to incorporate the full one-loop as well as dominant two-loop contributions to the effective scalar potential
and other important radiative corrections.
We will take advantage of this to go further in the study of no-scale models~\cite{Benhenni:2011yt}, 
implementing the minimization mechanism within SuSpect \cite{Djouadi:2002ze}. In a generalized no-scale inspired
framework, it first requires the definition of the soft parameters at the GUT scale
\be
B_0= b_0 \ m_{1/2},\;\; m_0= x_0 \ m_{1/2},\;\; A_0=a_0 \ m_{1/2}; 
\label{nsbc}
\ee
where the gaugino mass is the unique scale parameter, and the strict no-scale corresponds to 
$b_0=x_0=a_0=0$.   
Notice that the usual $\tan \beta$ input is replaced by $B_0$, the former 
being consistently derived at the electroweak scale.
In addition to this usual parameter set, we will have to consider a new one, 
in the form of a boundary condition $\eta_0$ for a vacuum energy term, following \cite{Kounnas:1994fr}.
\section{Renormalization Group invariant effective potential}
The vacuum energy term $\eta_0$ finds its roots in renormalization group (RG) invariance properties 
of the effective potential~\cite{VRGinv}.  
Adding one-loop contributions to the tree-level potential already ensures a more stable physical spectrum, 
but without the vacuum energy contribution the effective potential is not RG-invariant. 
In particular in the no-scale approach one is interested in the overall shape of the potential, and
obviously a meaningful minimum is  expected to be scale-independent, as much as possible 
perturbatively. The (one-loop) RG-invariant potential reads:
\be
V_{full} \equiv V_{tree}(Q) +V_{1-loop}(Q) +\tilde{\eta}(Q) m^4_{1/2}
\label{Vfull}
\ee
where as usual the one-loop contribution is expressed in terms of (field dependent) eigenmasses as
\be
V_{1-loop}(Q) = \frac {1}{64\pi ^2}
\sum_{all n} (-1)^{2n}  M_n^4(H_u,H_d)(\ln  \frac{ M_n^2(H_u,H_d)}{Q^2}-\frac 32)
\ee
and in Eq.~(\ref{Vfull}) the vacuum energy term is conveniently scaled by $m_{1/2}$ 
without much loss of generality. 
$\eta(Q)$ runs from $\eta_0$ at GUT scale to $\eta_{EW}$ at EW scale. We have checked that 
this term is not only crucial for RG-invariance and stability of the potential and corresponding 
physical minimum, but its contribution is also strongly correlated with the position of the minimum, 
and the corresponding value of the gaugino mass $\mhalf$.
\section{The minimization procedure}
The generalized no-scale electroweak minimization implies, in addition to the two usual EW minimizations  
$ \frac{\partial V_{full}}{\partial v_i}=0$, $i=u,d$, an extra minimization in the gaugino direction: 
$\frac{\partial V_{full}}{\partial m_{1/2}}  = 0 $. This will dynamically determine the soft 
parameters if all related to $m_{1/2}$ as in Eq.~(\ref{nsbc}). 
Assuming furthermore $\mu \sim m_{1/2}$, the latter minimization takes the convenient form~\cite{Kounnas:1994fr,Benhenni:2011yt} 
\be
V_{full}(m_{1/2}) +\frac{1}{128\pi^2} \sum_n (-1)^{2n} M^4_n(m_{1/2}) 
+ \frac{1}{4} m_{1/2}^5 \frac{d \tilde{\eta}_0}{d m_{1/2}} =0
\label{dVm12}
\ee 
where the last term is non-vanishing only in case $\tilde \eta_0$ may be a 
non-trivial function of $m_{1/2}$.
Upon minimizing the potential, care is to be taken when handling some of the physical constraints. 
Typically the right $m_Z$ mass constraint, $m^2_Z ={ v^2 \over 2 } (g'^2+g^2)$, 
and the pole-to-running mass relations
(mostly in the top quark sector due to strong dependence on the top quark Yukawa coupling),
$m^{pole}_{top} =  Y_t(Q) v_u(Q) (1+\delta^{RC}_y(Q)+\cdots) \; ,
\label{Ythresh}$
should be imposed only {\em after} the global minimum in the three directions $v_u, v_d, m_{1/2}$ has been 
found. This implies deviations of a few percent
in the $m_{1/2}$ minima values when taken into account properly~\cite{Benhenni:2011yt}.
\section{No-scale favored regions}
On phenomenological grounds, the no-scale mechanism generally favours a charged (mostly $\tilde \tau$) LSP
for $m_0=0$ or small enough,. 
This is not a problem as it is natural to consider the gravitino 
as the true LSP within this framework. Current sparticle mass limits from the 
LHC~\cite{LHClimits}  exclude small 
$\mhalf \lsim 300-350$ GeV values. Other indirect constraints, such as $B\to s\gamma$ measurements,
LEP Higgs mass bounds, etc, can be accomodated for sufficiently large $m_{1/2}$.
In our case, $m_{1/2}$ limits translate into bounds on $\eta_0$ values, favouring lower values $\eta_0\lsim 8-10$
(depending on other parameters, $B_0$ etc)~\cite{Benhenni:2011yt}. But it is 
still possible to have viable parameter regions with non-trivial $m_{1/2}$ minima, including
even a decoupled supersymmetric spectrum with a light SM-like Higgs, when $\eta_0 \simeq 0$.

\section{Gravitino dark-matter}
 For scenarios with a gravitino LSP (with stau as NLSP), all supersymmetric particles decay to the
 NLSP well before the latter has decayed to a gravitino, because all interactions to the gravitino 
are suppressed by the Planck mass. We first compute, using micrOMEGAs 2.0 \cite{Belanger:2006is}, 
the relic density $\Omega_{\rm NLSP}h^2$ the NLSP would have if it did not decay to the gravitino. Then 
assuming that each NLSP with mass $m_{\rm NLSP}$ decays to one gravitino, leads to the non-thermal 
contribution to the gravitino relic density
\be
\Omega_{3/2}^{\rm NTP}h^2={m_{3/2} \over m_{\rm NLSP}} \Omega_{\rm NLSP}h^2
\label{omntp}
\ee
with $h=0.73^{+0.04}_{-0.03}$ the Hubble constant.
\begin{figure}[h!]
\centering
   \includegraphics[width=6cm]{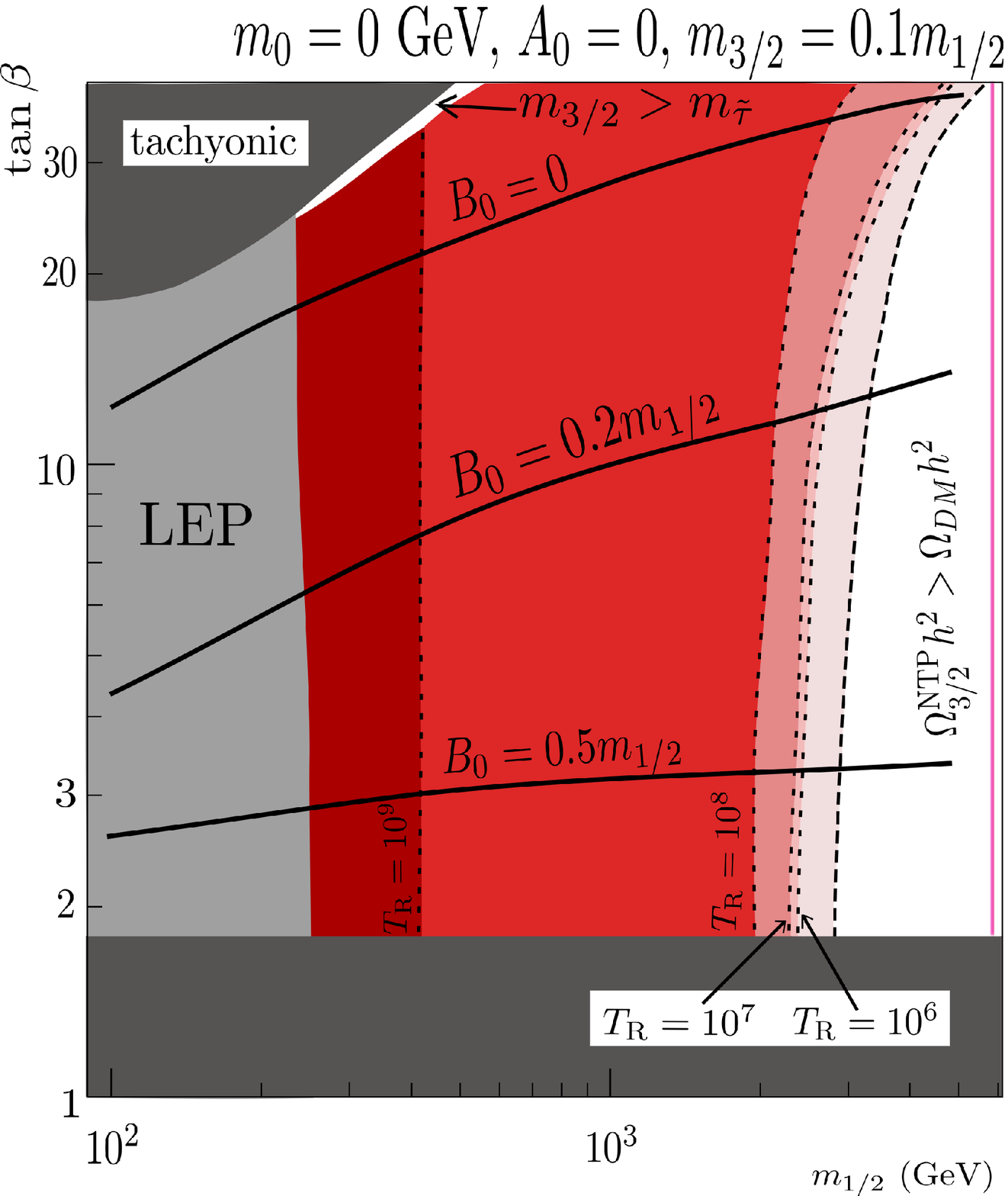}
\includegraphics[width=6cm]{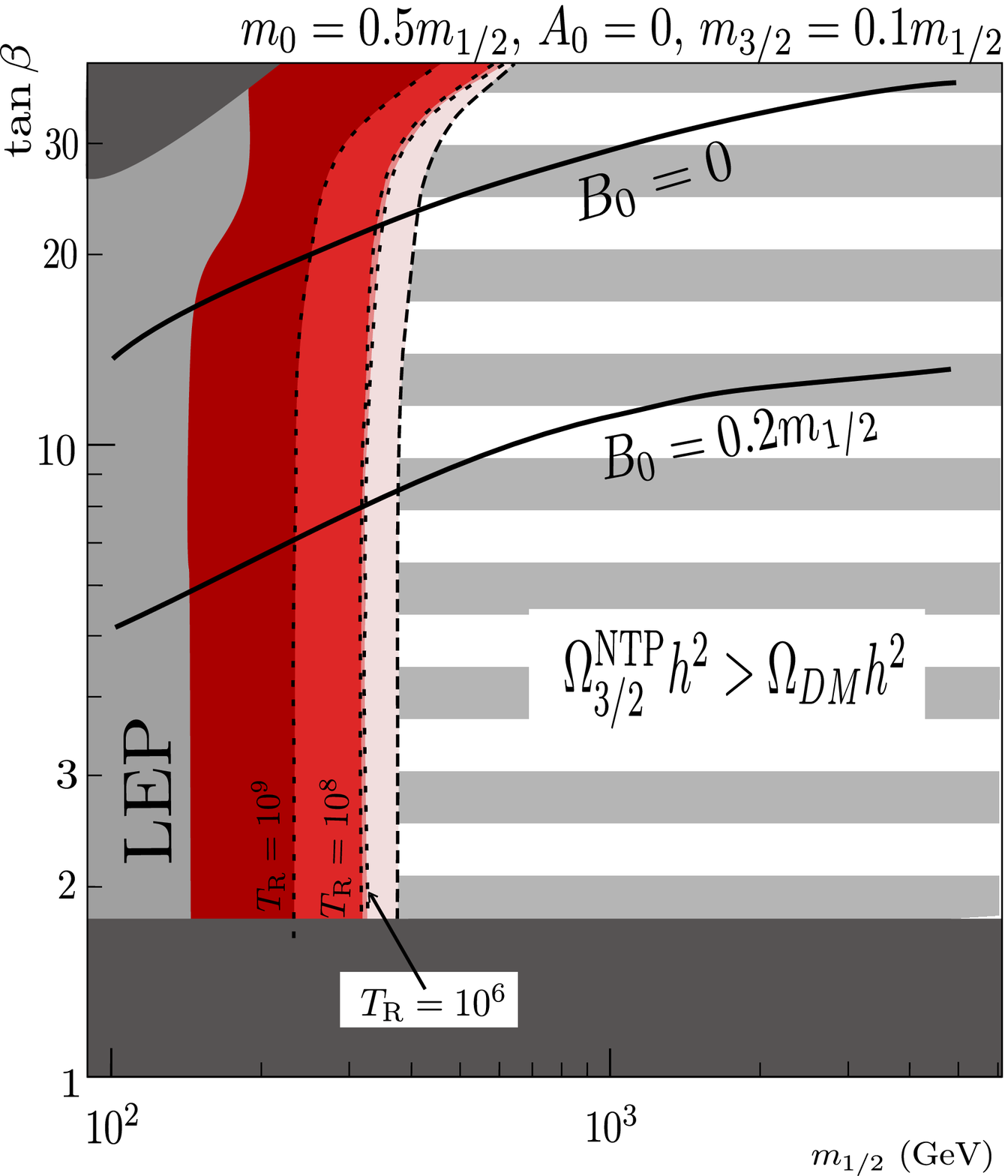}
 \caption{Gravitino LSP and relic density}
\end{figure}
The gravitino can also be produced during reheating after inflation. 
The gravitino relic 
density from such thermal production, $\Omega_{3/2}^{\rm TP}h^2$, is essentially 
controlled by the reheat temperature $T_R$ (see e.g \cite{Pradler:2006qh}).
Comparing the total gravitino relic density $\Omega_{3/2}^{\rm TP}h^2
+\Omega_{3/2}^{\rm NTP}h^2$,  
to WMAP constraints\cite{Spergel:2006hy}, will constrain $T_R$ together with the no-scale parameter space.
We illustrate two representative cases in the 
($\tan \beta$, $\mhalf$) plane, one for the strict no-scale scenario ($m_0=A_0=0$), and another less 
stringent scenario where the neutralino has some room as the LSP (though only for rather low $\mhalf$). 
One recovers consistency with the WMAP relic density constraint in a large part 
of the parameter space, provided that $T_R$ is sufficiently large, $T_R\gsim 10^6$ GeV. 
In particular even for the strict no-scale model $B_0=m_0=A_0=0$ 
there is a range for $m_{1/2}\sim 400-800$ GeV, $\tan\beta\sim 20-25$ compatible
with all present constraints, provided that the reheat temperature is 
$10^8-10^9$ GeV. 

\end{document}